\documentclass[reprint,amsmath,amssymb,apr,aip,onecolumn,nofootinbib]{revtex4-1}
\usepackage{graphicx}
\usepackage{graphics}
\usepackage{mathptmx}
\usepackage{times}
\usepackage{amsmath}
\usepackage{amssymb}
\usepackage{dcolumn}
\usepackage{bm}
\usepackage{units}
\usepackage{color}
\usepackage{soul}
\usepackage{multirow}
\graphicspath{{Figures/}}

\begin{document}
\title{Accurate deep neural network inference using computational phase-change memory}
\author{Vinay Joshi}\affiliation{IBM Research - Zurich, S\"{a}umerstrasse 4, 8803 R\"{u}schlikon, Switzerland}\affiliation{King's College London, Strand, London WC2R 2LS, United Kingdom}
\author{Manuel Le Gallo}\email{anu@zurich.ibm.com}\affiliation{IBM Research - Zurich, S\"{a}umerstrasse 4, 8803 R\"{u}schlikon, Switzerland}
\author{Simon Haefeli}\affiliation{IBM Research - Zurich, S\"{a}umerstrasse 4, 8803 R\"{u}schlikon, Switzerland}\affiliation{ETH Zurich, R\"{a}mistrasse 101, 8092 Zurich, Switzerland}
\author{Irem Boybat}\affiliation{IBM Research - Zurich, S\"{a}umerstrasse 4, 8803 R\"{u}schlikon, Switzerland}\affiliation{Ecole Polytechnique Federale de Lausanne (EPFL), 1015 Lausanne, Switzerland}
\author{S.R. Nandakumar}\affiliation{IBM Research - Zurich, S\"{a}umerstrasse 4, 8803 R\"{u}schlikon, Switzerland}
\author{Christophe Piveteau}\affiliation{IBM Research - Zurich, S\"{a}umerstrasse 4, 8803 R\"{u}schlikon, Switzerland}\affiliation{ETH Zurich, R\"{a}mistrasse 101, 8092 Zurich, Switzerland}
\author{Martino Dazzi}\affiliation{IBM Research - Zurich, S\"{a}umerstrasse 4, 8803 R\"{u}schlikon, Switzerland}\affiliation{ETH Zurich, R\"{a}mistrasse 101, 8092 Zurich, Switzerland}
\author{Bipin Rajendran}\affiliation{King's College London, Strand, London WC2R 2LS, United Kingdom}
\author{Abu Sebastian}\email{ase@zurich.ibm.com}\affiliation{IBM Research - Zurich, S\"{a}umerstrasse 4, 8803 R\"{u}schlikon, Switzerland}
\author{Evangelos Eleftheriou}\affiliation{IBM Research - Zurich, S\"{a}umerstrasse 4, 8803 R\"{u}schlikon, Switzerland}
\date{\today}

\begin{abstract}
In-memory computing is a promising non-von Neumann approach for making energy-efficient deep learning inference hardware. Crossbar arrays of resistive memory devices can be used to encode the network weights and perform efficient analog matrix-vector multiplications without intermediate movements of data. However, due to device variability and noise, the network needs to be trained in a specific way so that transferring the digitally trained weights to the analog resistive memory devices will not result in significant loss of accuracy. Here, we introduce a methodology to train ResNet-type convolutional neural networks that results in no appreciable accuracy loss when transferring weights to in-memory computing hardware based on phase-change memory (PCM). We also propose a compensation technique that exploits the batch normalization parameters to improve the accuracy retention over time. We achieve a classification accuracy of 93.7\% on the CIFAR-10 dataset and a top-1 accuracy on the ImageNet benchmark of 71.6\% after mapping the trained weights to PCM. Our hardware results on CIFAR-10 with ResNet-32 demonstrate an accuracy above 93.5\% retained over a one day period, where each of the 361,722 synaptic weights of the network is programmed on just two PCM devices organized in a differential configuration. 
\end{abstract}
\maketitle
\thispagestyle{myheadings}

\section{Introduction}\label{sec:intro}
Deep neural networks (DNNs) have revolutionized the field of artificial intelligence and have achieved unprecedented success in cognitive tasks such as image and speech recognition. Platforms for deploying the trained model of such networks and performing inference in an energy-efficient manner are highly attractive for edge computing applications. In particular, internet-of-things battery-powered devices and autonomous cars could especially benefit from fast, low-power, and reliably accurate DNN inference engines. Significant progress in this direction has been made with the introduction of specialized hardware for inference operating at reduced digital precision (4 to 8-bit), such as Google's tensor processing unit (TPU) \cite{jouppi2017} and low-power graphical processing units (GPUs) such as NVIDIA T4 \cite{jia2019}. While these platforms are very flexible, they are based on architectures where there is a physical separation between memory and processing units. The models are typically stored in off-chip memory, leading to constant shuttling of data between memory and processing units, which limits the maximum achievable energy efficiency.


In order to reduce the data transfers to a minimum in inference accelerators, a promising avenue is to employ in-memory computing using non-volatile memory devices \cite{shafiee2016}. Both charge-based storage devices, such as Flash memory \cite{merrikh2017}, and resistance-based (memristive) storage devices, such as metal-oxide resistive random-access memory (ReRAM) \cite{chen2019,hu2018} and phase-change memory (PCM) \cite{Y2018legalloNatureElectronics,Boybat2018,Y2018ambrogioNature} are being investigated for this. In this approach, the network weights are encoded as the analog charge state or conductance state of these devices organized in crossbar arrays, and the matrix-vector multiplications during inference can be performed in-situ in a single time step by exploiting Kirchhoff's circuit laws. The fact that these devices are non-volatile (the weights will be retained when the power supply is turned off) and have multi-level storage capability (a single device can encode an analog range of values as opposed to 1 bit) is very attractive for inference applications. However, due to the analog nature of the weights programmed in these devices, only limited precision can be achieved in the matrix-vector multiplications and this could limit the achievable inference accuracy of the accelerator.

One potential solution to this problem is to train the network fully on hardware \cite{Y2018nandakumarISCAS,Y2018ambrogioNature}, such that all hardware non-idealities would be de facto included as constraints during training. Another similar approach is to perform partial optimizations of the hardware weights after transferring a trained model to the chip \cite{mohanty2017,gonugondla2018}. The drawback of these approaches is that every neural network would have to be trained on each individual chip before deployment. Off-line variation-aware training schemes have also been proposed, where hardware non-idealities such as device-to-device variations \cite{liu2015,chen2017}, defective devices \cite{chen2017}, or IR drop \cite{liu2015} are first characterized and then fed into the training algorithm running in software. However, these approaches would require characterizing and training the neural network from scratch for every chip. A more practical approach would be to have a single custom generic training algorithm that is run entirely in software which would make the network immune to most of the hardware non-idealities, but at the same time would require only very little knowledge about the specific hardware it will be deployed on. In this way, the model would have to be trained only once and could be deployed on a multitude of different chips. To this end, several works have proposed to inject noise in the training algorithm to the layer inputs \cite{Y2019moonTVLSI}, synaptic weights \cite{Y2017miyashitaJSSC}, and pre-activations \cite{klachko2019,Rekhi2019}. However, previous demonstrations have generally been limited to rather simple and shallow networks, and experimental validations of the effectiveness of the various approaches have been missing. We are aware of one recent work that analyzed more complex problems such as ImageNet classification \cite{Rekhi2019}, however the hardware model used was  rather abstract and no experimental validation was presented.

In this work, we explore injecting noise to the synaptic weights during the training of DNNs in software as a generic method to improve the network resilience against analog in-memory computing hardware non-idealities. We focus on the ResNet convolutional neural network (CNN) architecture, and introduce a number of techniques that allow us to achieve a classification accuracy of 93.7\% on the CIFAR-10 dataset and a top-1 accuracy of 71.6\% on the ImageNet benchmark after mapping the trained weights to PCM synapses. In contrast to previous approaches, the noise injected during training is crudely estimated from a one-time all-around hardware characterization, and captures the combined effect of read and write noise without introducing additional noise-related training hyperparameters. We validate the training approach through hardware/software experiments, where each of the 361,722 weights of ResNet-32 is programmed on two PCM devices of a prototype chip, and the rest of the network functionality is simulated in software. We achieve an experimental accuracy of $93.75\%$ after programming, which stays above $92.6\%$ over a period of 1 day. To improve the accuracy retention further, we develop a method to periodically calibrate the batch normalization parameters to correct the activation distributions during inference. We demonstrate a significant improvement in the accuracy retention with this method (up to $93.5\%$ on hardware for CIFAR-10) compared with a simple global scaling of the layers' outputs, at the cost of additional digital computations during calibration. Finally, we discuss our training approach with respect to other methods and quantify the tradeoffs in terms of accuracy and ease of training. 

\section{Results}
\subsection{Problem statement}\label{sec:problem}

\begin{figure*}
\includegraphics[scale=0.95]{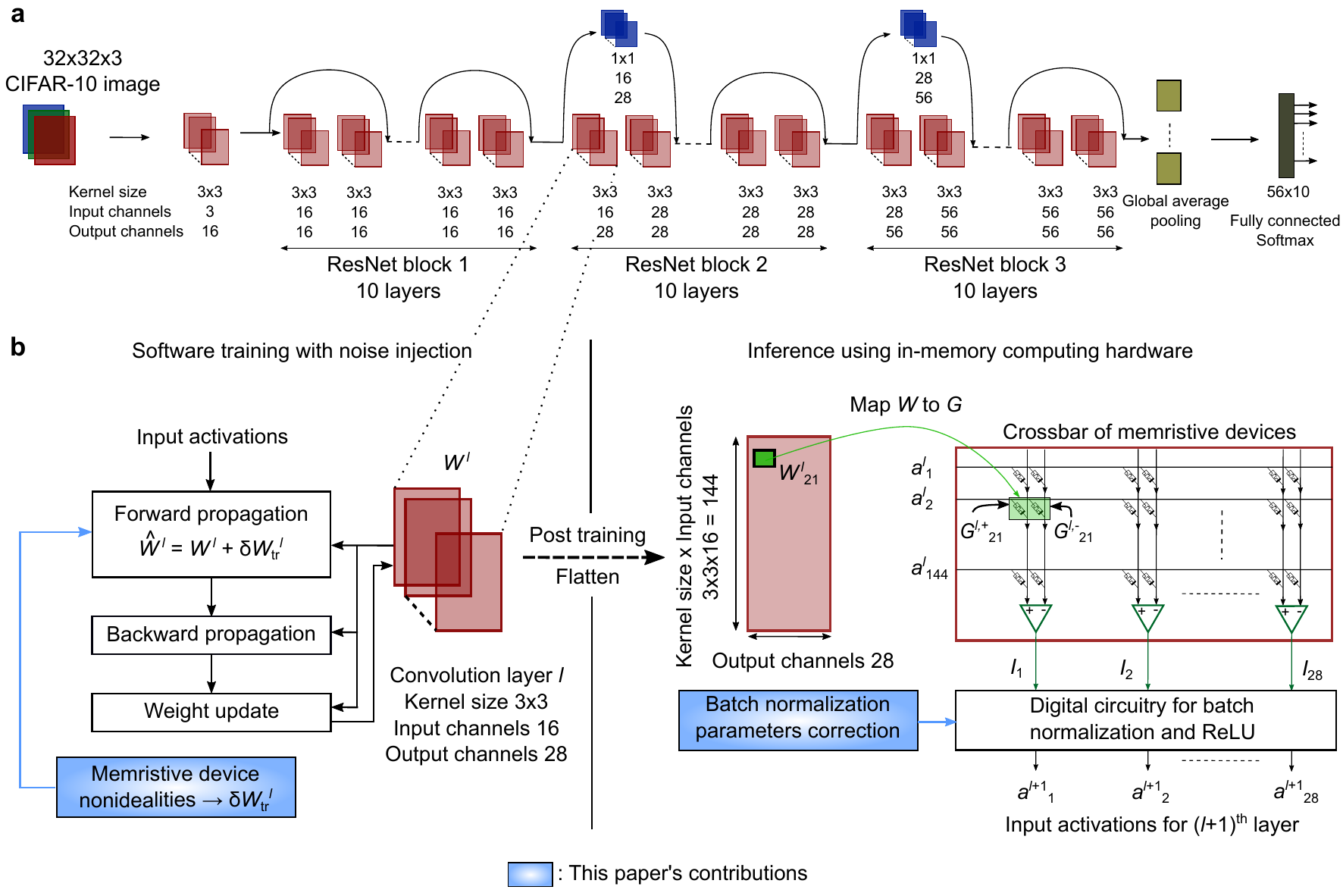}
\caption{\textbf{a}, ResNet-32 network architecture for CIFAR-10 image classification. The architecture of ResNet-32 used in this study is a slightly modified version of the original implementation \cite{Y2016heCVPR} with fewer input and output channels in ResNet blocks 2 and 3. \textbf{b}, Training and inference of an example layer of ResNet-32 according to the methodology proposed in this paper. Software training is performed by injecting a random noise term to the weights used during the forward propagation, $\delta W_{tr}^l$, which is representative of the combined read and write noise of the memristive devices used during inference (see Section \ref{sec:training}). When transferring the weights of a convolution layer to memristive crossbars for inference, they are flattened into a 2-D matrix by collapsing each filter into a single vector programmed on a crossbar column, and stacking all filters on separate columns \cite{Gokmen2017}. The weights are then programmed as the differential conductance of two memristive devices. Input activations $a^l$ are applied as voltages on the crossbar rows. The output current from the column containing the $G^{-}$ devices is subtracted from the one from the column containing the $G^{+}$ devices. The differential current output $I$ from the crossbar then routes to a digital circuitry that performs batch normalization and the corresponding rectified linear unit (ReLU) activation function, in order to obtain the input activations for the next layer $a^{l+1}$. The final softmax activation function can be performed off-chip if required. An optional correction of the batch normalization parameters can be periodically performed to improve the accuracy retention over time (see Section \ref{sec:experiment} and \ref{sec:AdaBS}). The input image is padded with zero values at the border to ensure that the convolution operation preserves the height and width of the image. Therefore, considering an input image of size $n \times n$, the convolution operation can be performed in $n^2$ matrix-vector multiplication cycles.
} \label{figure1}
\end{figure*}

For our experiments, we consider two residual networks on two different datasets: ResNet-32 on the CIFAR-10 dataset, and ResNet-34 on the ImageNet dataset \cite{Y2016heCVPR}. As shown in Fig.\ \ref{figure1}\textbf{a}, ResNet-32 consists of 3 different ResNet blocks with ten $3\times3$ kernels each, and is used to classify $32\times32$-pixel RGB images that belong to one out of 10 classes (see Methods). The network contains 361,722 synaptic weights. The ResNet-34 network used for the 1000-class ImageNet dataset is shown in Supplementary Fig.\ 1. The main differences compared to ResNet-32 are the number and size of the ResNet blocks, and a larger number of input/output channels (see Methods). 

The weights of all convolution layers along with the fully connected layer of ResNet-32 can be mapped on memristive crossbar arrays as explained in Fig.\ \ref{figure1}\textbf{b}. Each synaptic weight can be mapped on a differential pair of memristive devices that are located on two different columns. For a given layer $l$, the synaptic weight $W_{ij}^{l}$ of the $(i,j)^{th}$ synaptic element is represented by the \textit{effective} synaptic conductance $G_{ij}^{l}$ given by
\begin{equation}\label{synaptic_cond}
G_{ij}^{l} = G_{ij}^{l,+} - G_{ij}^{l,-},
\end{equation}
where $G_{ij}^{l,+} $ and $ G_{ij}^{l,-} $ are the conductance values of the two devices forming the differential pair. Those device conductance values are defined as the effective conductance perceived in the operation of a non-ideal memristive crossbar array, and therefore include all the circuit non-idealities from the crossbar and peripheral circuitry. 

The mapping between the synaptic weight $W_{ij}^{l}$ obtained after software training and the corresponding synaptic conductance is given by
\begin{equation}\label{synaptic_map}
G_{ij}^{l} = W_{ij}^{l} \times \frac{G_{max}}{W_{max}^{l}} + \delta G_{ij}^{l} = G_{T,ij}^{l} + \delta G_{ij}^{l},
\end{equation}
where $G_{max}$ is the maximum reliably programmable device conductance and $W_{max}^{l}$ is the maximum absolute synaptic weight value of layer $l$. $\delta G_{ij}^{l}$ represents the synaptic conductance error from the ideal target conductance value $G_{T,ij}^{l} = W_{ij}^{l} \times \frac{G_{max}}{W_{max}^{l}}$. $\delta G_{ij}^{l}$ is a time-varying random variable that describes the effects of non-ideal device programming (inaccuracies associated with write) and conductance fluctuations over time (inaccuracies associated with read).  Possible factors leading to such conductance errors include inaccuracies in programming the synaptic conductance to $G_{T,ij}^{l}$, $1/f$ noise from memristive devices and circuits, temporal conductance drift, device-to-device variations, defective (stuck) devices, and circuit non-idealities (e.g. IR drop).

Clearly, a direct mapping of the synaptic weights of a DNN trained with 32-bit floating point (FP32) precision to the same DNN with memristive synapses is expected to degrade the network accuracy due to the added error in the weights arising from $\delta G_{ij}^{l}$. For existing memristive technologies, the magnitude of $\delta G_{ij}^{l}$ may range from $1-10$\% of the magnitude of $G_{T,ij}^{l}$ \cite{Y2018legalloNatureElectronics}, which in general is not tolerable by DNNs trained with FP32 without any constrains. Imposing such errors as constraints during training can be beneficial in improving the network accuracy. In fact, quantization of the weights or activations \cite{merolla2016}, and injecting noise on the weights \cite{blundell2015}, activations \cite{gulcehre2016} or gradients \cite{neelakantan2015} have been widely used as DNN regularizers during training to reduce overfitting on the training dataset \cite{an1996,Jim1994}. These techniques can improve the accuracy of DNN inference when it is performed with the same model precision as during training. However, achieving baseline accuracy while performing DNN inference on a model which is \textit{inevitably} different from the one obtained after training, as it is the case for any analog in-memory computing hardware, is a more difficult problem and requires additional investigations.

Although a large body of efficient techniques to train DNNs with reduced digital precision has been reported \cite{Y2015guptaICML,McKinstry2018}, it is unlikely that such procedures can generally be applied as-is to analog in-memory computing hardware due to the random nature of $\delta G_{ij}^{l}$. Since quantization errors coming from rounding to reduced fixed-point precision are not random, DNNs trained in this way are not \textit{a priori} expected to be suitable for deployment on analog in-memory computing hardware. 
Techniques that inject random Gaussian noise during training are a much more natural fit to make the network robust to errors from analog in-memory computing hardware. As early as in 1994, it was shown that injecting noise on the synaptic weights during training enhances the tolerance to weight perturbations of multi-layer perceptrons, and the application of this technique to analog neural hardware was discussed \cite{murray1994}. Recent works have also proposed to apply noise to the layer inputs or pre-activations in order to improve the network tolerance to hardware noise \cite{Y2019moonTVLSI,Rekhi2019}. 
In this work, we follow the original approach of Murray et al. \cite{murray1994} of injecting Gaussian noise to the synaptic weights during training. Next, we discuss different techniques that we employed together with synaptic weight noise in order to improve the accuracy of inference on ResNet and achieve close to software-equivalent accuracy after transferring the weights to PCM hardware.

\subsection{Training procedure}\label{sec:training}

When performing inference with analog in-memory computing hardware, the DNN experiences errors primarily due to (i) inaccurate programming of the network weights onto the devices (write noise) and (ii) temporal fluctuations of the hardware weights (read noise). We can cast the effect of these errors into a single error term $\delta G_{ij}^{l}$ that distorts each synaptic weight when performing forward propagation during inference. Hence, we propose to add random noise that corresponds to the error induced by $\delta G_{ij}^{l}$ to the synaptic weights at each forward pass during training (see Fig.\ \ref{figure1}\textbf{b}). 
The backward pass and weight updates are performed with weights that did not experience this noise. 
We found that adding noise to the weights only in the forward propagation is sufficient to achieve close to baseline accuracy for a noise magnitude comparable to that of our hardware, and adding noise during the backward propagation did not improve the results further. For simplicity, we assume that $\delta G_{ij}^{l}$ is Gaussian distributed, which is usually the case for analog memristive hardware. Weights are linearly mapped to the entire conductance range $G_{max}$ of the hardware, hence the standard deviation $\sigma_{\delta W_{tr}}^{l}$ of the Gaussian noise on weights to be applied during training, for a layer $l$, can be computed as
\begin{equation} \label{std_compute}
\frac{\sigma_{\delta W_{tr}}^{l}}{W_{max}^{l}} \equiv \eta_{tr} = \frac{\sigma_{\delta G}}{G_{max}},
\end{equation}
where $\sigma_{\delta G}$ is a representative standard deviation of the combined read and write noise measured from hardware. During training, the weight distribution of every layer and hence $W_{max}^{l}$ changes, therefore $\sigma_{\delta W_{tr}}^{l}$ is recomputed after every weight update so that $\eta_{tr}$ stays constant throughout training. We found this to be especially important in achieving good training convergence with this method.

Weight initialization can have a significant effect on DNN training \cite{he_init}. Two different weight initializations can lead to completely different minima when optimizing the network objective function. The network optimum when training with additive noise could be closer to the FP32 training optimum than to a completely random initialization. So it can be beneficial to initialize weights from a pretrained baseline network and then retrain this network by injecting noise. A similar observation was reported for training ResNet with reduced digital precision \cite{McKinstry2018}. For achieving high classification accuracy in our experiments, we found this strategy more helpful than random initialization.

The noise injected during training according to Eq.\ \eqref{std_compute} is closely related to the maximum weight of a layer, and can thus grow uncontrollably with outlier weight values. Controlling the weight distribution in a desirable range can improve the network training convergence and makes the mapping of weights to hardware with limited conductance range easier. We therefore clip the synaptic weights at layer $l$ after every weight update in the range $[-\alpha \times \sigma_{W^{l}},\alpha \times \sigma_{W^{l}}]$, where $\sigma_{W^{l}}$ is the standard deviation of weights in layer $l$ and $\alpha$ is a tunable hyperparameter. In our studies, $\alpha = 2.0$ and $\alpha = 2.5$ worked the best for ResNet-32 and ResNet-34, respectively.

DNN convergence accuracy, in general, is sensitive to the learning rate used during training. Since we initialize the network parameters from a baseline network, using the same learning rate scheduling as that of the baseline network does not guarantee accurate convergence. To choose appropriate learning rate scheduling for ResNet-32, we first forward propagate the training set on the pretrained baseline network with injected synaptic weight noise and note the resulting accuracy. We note the learning rate evolution starting from this accuracy in the baseline network training curve until convergence, and use the same learning rate evolution while retraining the network by injecting noise.

\begin{figure}
\includegraphics[scale=0.95]{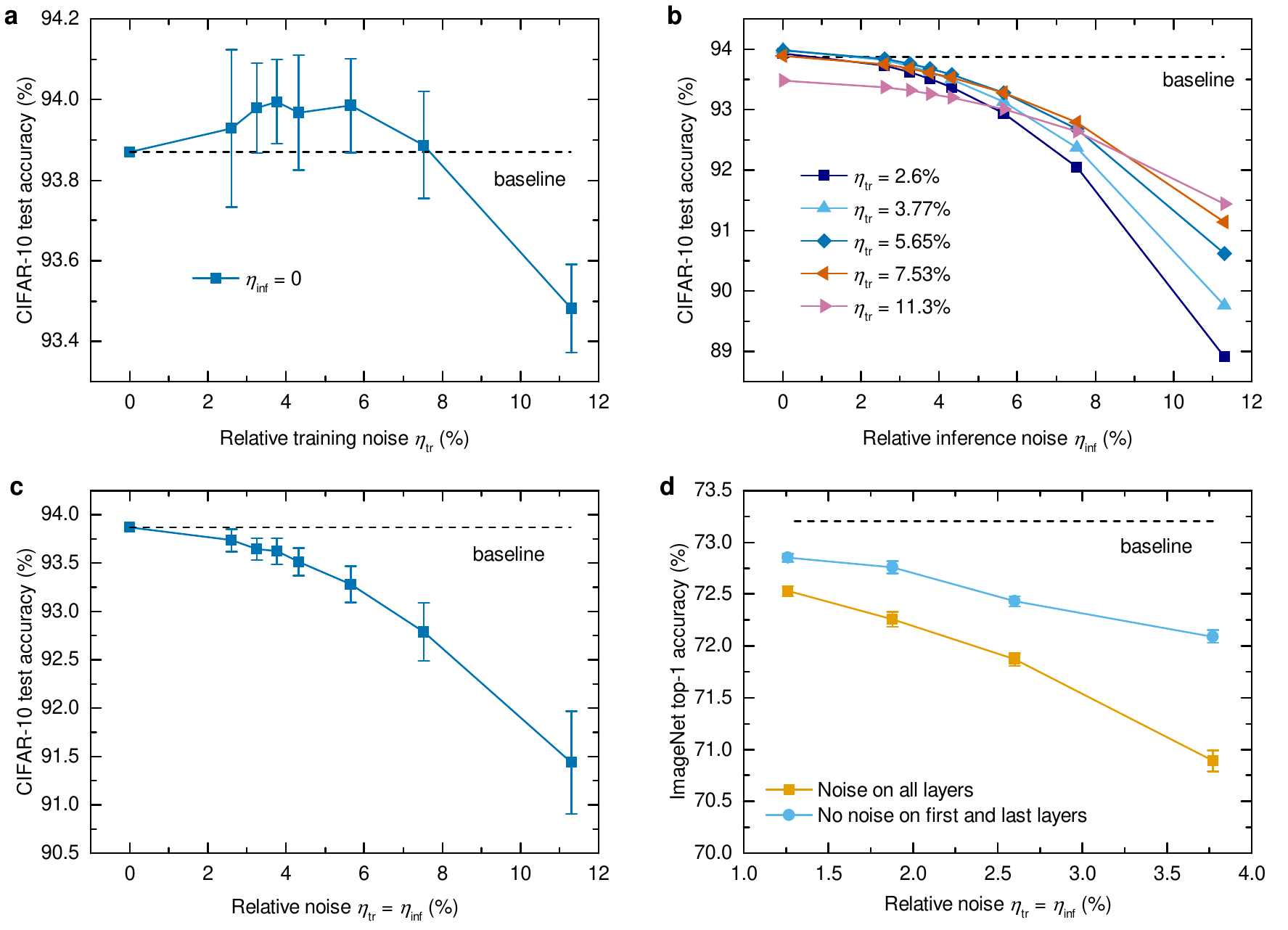}
\caption{\textbf{a}, Test accuracy on CIFAR-10 obtained for different amounts of relative injected weight noise during training $\eta_{tr}$ without inducing any perturbation during inference ($\eta_{inf}=0$). When $\eta_{tr} > 8\%$, the training convergence starts to become affected by the high noise and it is not possible anymore to reach the software baseline within the same number of epochs. The error bars represent the standard deviation over 10 training runs. \textbf{b}, Test accuracy on CIFAR-10 obtained for the networks trained with different amounts of relative weight noise $\eta_{tr}$ as a function of the weight noise injected during inference $\eta_{inf}$. In most cases, $\eta_{tr}$ can be increased above $\eta_{inf}$ up to a certain point and still lead to comparable or slightly higher (within $\approx 0.1\%$) test accuracy than for $\eta_{tr} = \eta_{inf}$. However, when $\eta_{tr}$ becomes much higher than $\eta_{inf}$, the test accuracy decreases due to the inability of the network to achieve baseline accuracy when $\eta_{tr} > 8\%$. Each data point represents the average over 10 training runs and 100 inference runs. \textbf{c}, Test accuracy on CIFAR-10 as a function of $\eta_{tr} = \eta_{inf}$. The error bars represent the standard deviation over 100 inference runs averaged over 10 training runs. \textbf{d}, Top-1 accuracy on ImageNet as a function of $\eta_{tr} = \eta_{inf}$. The error bars represent the standard deviation over 10 inference runs on a single training run. } 
\label{figure2}
\end{figure}

We performed simulations to characterize the inference performance after training incorporating the injection of Gaussian noise in conjunction with the techniques presented above. We computed the classification accuracy for different amounts of injected noise $\eta_{tr}$ during training. We also show how the accuracy is affected when the inference weights are perturbed by a certain amount of relative noise $\eta_{inf} \equiv \frac{\sigma_{\delta W_{inf}}^{l}}{W_{max}^{l}}$, where $\sigma_{\delta W_{inf}}^{l}$ is the standard deviation of the noise injected to the weights of layer $l$ before performing inference on the test dataset.

The test accuracy of ResNet-32 on CIFAR-10 obtained for different amounts of noise injected during training, without inducing any perturbation during inference ($\eta_{inf}=0$), is plotted in Fig.\ \ref{figure2}\textbf{a}. It can be seen that the training algorithm is able to achieve a test accuracy close to the software baseline of $93.87\%$ with up to approximately $\eta_{tr} = 8\%$. The tolerance of the networks trained with different amounts of $\eta_{tr}$ to weight perturbations during inference, $\eta_{inf}$, is shown in Fig.\ \ref{figure2}\textbf{b}. For a given value of $\eta_{inf}$, in general, the highest test accuracy can be obtained for the network that has been trained with a comparable amount of synaptic weight noise, i.e. for $\eta_{tr} \approx \eta_{inf}$. The test accuracy for $\eta_{tr} = \eta_{inf}$ is shown in Fig.\ \ref{figure2}\textbf{c}. It can be seen that for up to $\eta_{inf} = 5\%$, an accuracy within $0.5\%$ of the software baseline is achievable. The impact of the weight initialization, clipping, and learning rate scheduling on the accuracy is shown in Supplementary Fig.\ 2. Not incorporating any of those three techniques results in at least 1\% drop in test accuracy for $\eta_{tr} = \eta_{inf} = 3.8\%$. 

The top-1 accuracy of ResNet-34 on ImageNet for $\eta_{tr} = \eta_{inf}$ is shown in Fig.\ \ref{figure2}\textbf{d}. 
Consistent with previous observations \cite{Rekhi2019,McKinstry2018}, we found that the network recovers high accuracy extremely quickly when retraining with additive noise due to quick updates of the batch normalization parameters (see Supplementary Note 1), and obtained satisfactory convergence after only 8 epochs. 
The accuracy on ImageNet is much more sensitive to the noise injected during training than for CIFAR-10, and when noise in injected on all layers, there is more than $0.5\%$ accuracy drop from the baseline even down to $\sim 1.2\%$ relative noise. In the literature, many network compression techniques allow higher precision for the first and last layers, which are more sensitive to noise \cite{rastegari2016,McKinstry2018}. We applied the same simplification to our problem, which means that we removed the noise during training on the first convolutional layer and the last dense layer, and performed inference with the first and last layer without noise. The obtained accuracy after training, by injecting the same training and inference noise as previously, can be increased by more than 1\% with this technique (see Fig.\ \ref{figure2}\textbf{d}). 

\subsection{Weight transfer to PCM-based synapses}\label{sec:weightTransfer}
\begin{figure}
\includegraphics[scale=0.95]{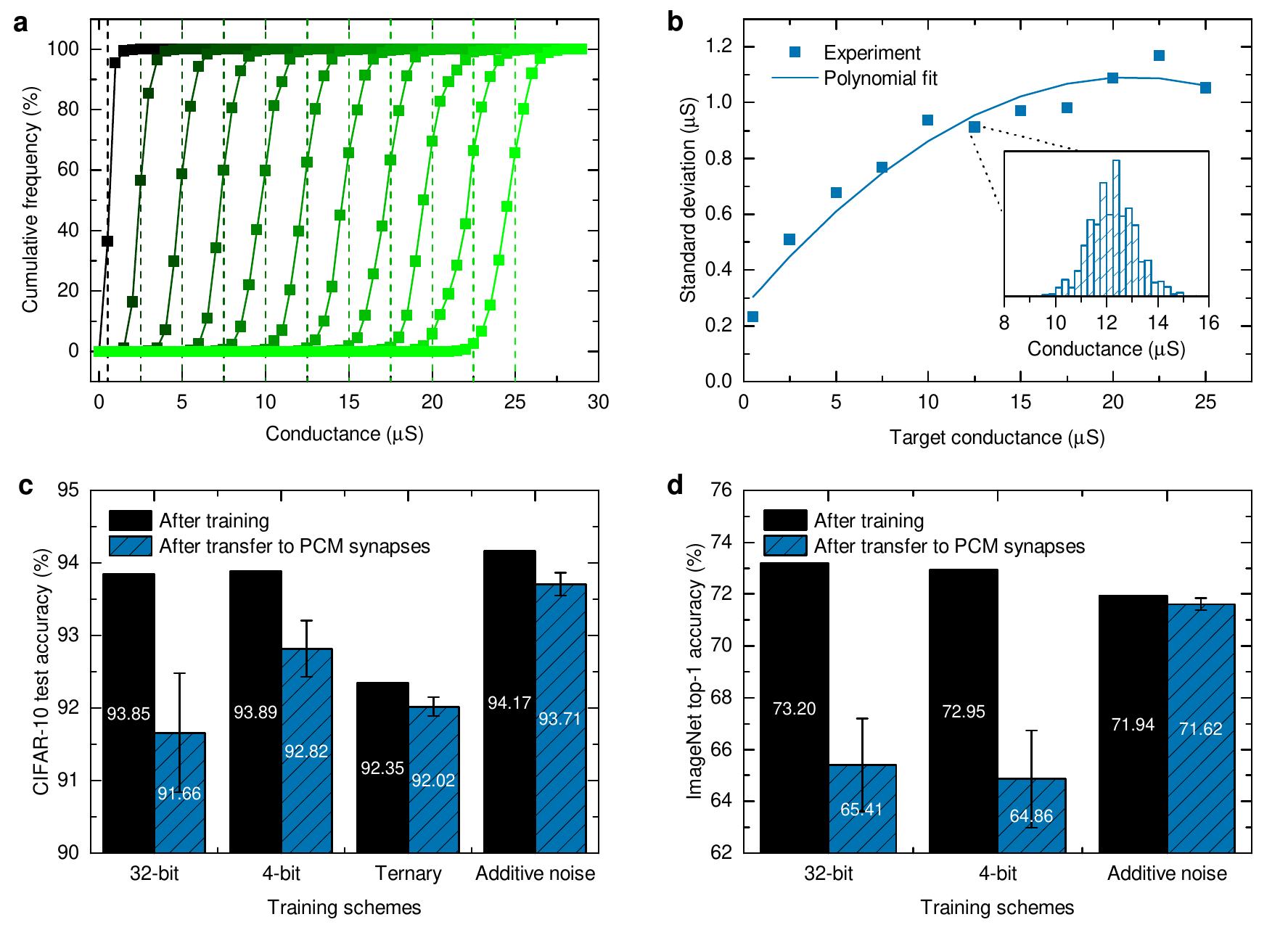}
\caption{\textbf{a}, Cumulative distributions of 11 representative iteratively programmed conductance levels on 10,000 PCM devices per level. The vertical dashed lines denote the target conductance for each level. \textbf{b}, Conductance standard deviation of the 11 levels as a function of target conductance. The inset shows a representative conductance distribution of one level. \textbf{c}, Test accuracy on CIFAR-10 after software training and after weight transfer to PCM synapses for different training schemes. \textbf{d}, Top-1 accuracy on ImageNet after software training and after weight transfer to PCM synapses for different training schemes. In \textbf{c} and \textbf{d}, the error bars represent the standard deviation over 10 inference runs. } \label{figure3}
\end{figure}

In order to experimentally validate the effectiveness of the above training methodology, we performed experiments on a prototype multi-level PCM chip comprising 1 million PCM devices fabricated in \unit[90]{nm} CMOS baseline technology \cite{Y2010closeIEDM}. PCM is a memristive technology which records data in a nanometric volume of phase-change material sandwiched between two electrodes \cite{Y2016burrJETCAS}. The phase-change material is in the low-resistive crystalline phase in an as-fabricated device. By applying a current pulse of sufficient amplitude (typically referred to as the RESET pulse) an amorphous region around the narrow bottom electrode is created via a melt-quench process. The device will be in a low conductance state if the high-resistive amorphous region blocks the current path between the two electrodes. The size of the amorphous region can be modulated in an almost completely analog manner by the application of suitable electrical pulses. Hence, a continuum of conductance values can be programmed in a single PCM device over a range of more than two orders of magnitude.

An optimized iterative programming algorithm was developed to program the conductance values in the PCM devices with high accuracy (see Methods). The experimental cumulative distributions of conductance values for 11 representative programmed levels, measured approximately 25 seconds after programming, are shown in Fig.\ \ref{figure3}\textbf{a}. The standard deviation of these distributions is extracted and fitted with a polynomial function of the target conductance (dashed lines in Fig.\ \ref{figure3}\textbf{a}) as shown in Fig.\ \ref{figure3}\textbf{b}. For all levels, we  achieve a standard deviation less than \unit[1.2]{$\mu$S}, which is more than 2 times lower than that reported in previous works on nanoscale  PCM arrays for a similar conductance range \cite{legalloTED2018,tsai2019}. 

To study the effect of weight transfer to PCM synapses, Eq.\ \eqref{synaptic_map} is computed using the conductance standard deviation measured from hardware. $\delta G_{ij}^{l}$ is modeled as a Gaussian distributed random variable with 0 mean and standard deviation given by the fitted curve of Fig.\ \ref{figure3}\textbf{b} for the corresponding target conductance, $\vert G_{T,ij}^{l} \vert$, computed with \unit[$G_{max} = 25$]{$\mu$S}. The resulting test accuracy obtained after software training and after weight transfer to PCM synapses for ResNet-32 on CIFAR-10 is shown in Fig.\ \ref{figure3}\textbf{c} for different training procedures. It can be seen that standard FP32 training without constraints performs the worst after transfer to PCM synapses. Training with 4-bit precision weights (using the method described in Ref. \onlinecite{McKinstry2018}), which is roughly the effective precision of our PCM devices \cite{legalloTED2018}, improves the performance after transfer with respect to FP32, but nevertheless the accuracy decreases by more than 1\% after transferring the 4-bit weights to PCM. Training ternary digital weights \cite{li2016ternary} leads to a lower performance drop ($< 0.5\%$) when transferring weights to PCM, although we were not able to reach the FP32 baseline with ternary weights on this network. Therefore the accuracy after transfer is worse than for the 4-bit weights. When performing training by injecting Gaussian noise as described in Section \ref{sec:training} with $\eta_{tr} = 3.8\%$, corresponding to \unit[$\sigma_{\delta G} = 0.94$]{$\mu$S} (median of the 11 values reported in Fig.\ \ref{figure3}\textbf{b}), the best overall performance after transfer to PCM is obtained. The resulting accuracy of 93.7\% is less than $0.2\%$ below the FP32 baseline. A rather broad range of values of $\eta_{tr}$ lead to a similar resulting accuracy (see Supplementary Fig.\ 3), demonstrating that $\eta_{tr}$ does not have to be very precisely determined for obtaining satisfactory results on PCM. The accuracy obtained without perturbing the weights after training by injecting noise is slightly higher than the FP32 baseline, which could be attributed to improved generalization resulting from the additive noise training.

The top-1 accuracy for ResNet-34 on ImageNet after transfer to PCM synapses for different training procedures is shown in Fig.\ \ref{figure3}\textbf{d}. Training with additive noise increases the accuracy by approximately 6\% on PCM compared with FP32 and 4-bit\cite{McKinstry2018} training. The accuracy of $71.6\%$ achieved with additive noise training on PCM is significantly higher than that reported in Fig.\ \ref{figure2}\textbf{d} with $\eta_{inf} = 3.8\%$, which could be attributed to a high percentage of network weights mapped to low conductance values with lower standard deviation than the median of \unit[$0.94$]{$\mu$S}. 

\subsection{Hardware/software inference experiment on CIFAR-10}\label{sec:experiment}
Although we could achieve good test accuracy after weight transfer to PCM synapses as shown in the previous section, an important challenge for any analog in-memory computing hardware is to be able to retain this accuracy over time. This is especially true for PCM due to the high $1/f$ noise experienced in these devices as well as temporal conductance drift. The conductance values in PCM drift over time $t$ according to the relation $G(t) = G(t_0)(t/t_0)^{-\nu}$, where $G(t_0)$ is the conductance measured at time $t_0$ after programming and $\nu$ is the drift exponent, which depends on the device, phase-change material, and phase configuration of the PCM ($\nu$ is higher for the amorphous than the crystalline phase) \cite{Y2018legalloAEM}. In our PCM devices, $\nu \approx 0.06$ on average. Therefore, it is essential to measure experimentally how the test accuracy evolves over time during inference with PCM.

\begin{figure*}
\includegraphics[scale=0.95]{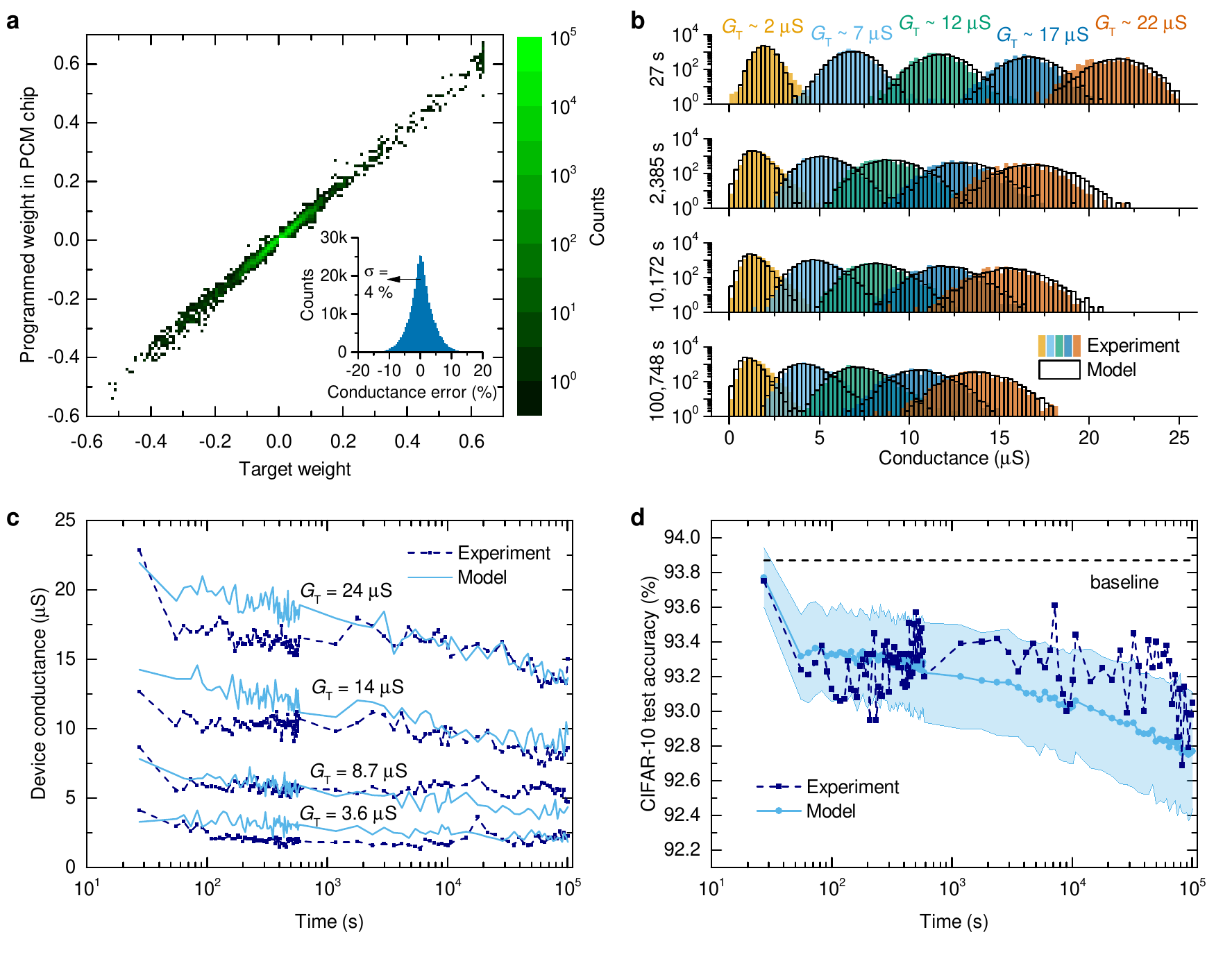}
\caption{\textbf{a}, Scatter plot of weights programmed in the PCM chip versus target weights obtained after training. The inset shows the distribution of the relative error between programmed and target synaptic conductance, $(G_{ij}^{l} - G_{\mathrm{T},ij}^{l})/G_\mathrm{max}$, and its standard deviation $\sigma$. \textbf{b}, Distributions of programmed devices whose target conductance fall within \unit[$0.25$]{$\mu$S} from 5 representative $G_T$ values. The distributions are shown at four different times spanning the experiment duration of one day. The filled bars are the measured hardware data, the black lines are the PCM model. \textbf{c}, Individual device conductance evolution over time of four arbitrarily picked devices from the chip programmed to four distinct $G_T$ values, along with one PCM model realization for the same $G_T$ values. \textbf{d}, Measured test accuracy over time from the weights of the PCM chip, along with the corresponding PCM model match. A global drift compensation (GDC) procedure is performed for every layer before performing inference on the test set. The filled areas from the PCM model correspond to one standard deviation over 25 inference runs. } \label{figure4}
\end{figure*}

Here, we present experiments where all 361,722 synaptic weights of ResNet-32 trained with $\eta_{tr} = 3.8\%$ are programmed individually on two PCM devices of the chip. Depending on the sign of $G_{T,ij}^{l}$, either $G_{ij}^{l,+}$ or $G_{ij}^{l,-}$ is iteratively programmed to $\vert G_{T,ij}^{l} \vert$, and the other device is RESET close to \unit[0]{$\mu$S} with a single pulse of \unit[450]{$\mu$A} amplitude and  \unit[50]{ns} width. The iterative programming algorithm converged on $99.1\%$ of the devices programmed to nonzero conductance, and no screening for defective devices on the chip was performed prior to the experiments. The scatter plot of the PCM weights measured approximately 25 seconds after programming versus the target weights $W_{ij}^{l}$ is shown in Fig.\ \ref{figure4}\textbf{a}. After programming, the PCM analog conductance values were periodically read from hardware over a period of 1 day, scaled to the network weights, and reported to the software that performed inference on the test dataset (see Methods).

In addition to the experiment, we developed an advanced behavioral model of the hardware in order to precisely capture the conductance evolution over time during inference (see Supplementary Note 2). The model is built based on an extensive experimental characterization of the array-level statistics of hardware noise and drift. Conductance drift is modeled using a Gaussian distributed drift exponent across devices, whose mean and standard deviation both depend on the target conductance state $\vert G_{T,ij}^{l} \vert$. Conductance noise with the experimentally observed $1/f^{1.21}$ frequency dependence is also incorporated with a magnitude that depends on the target conductance state and time. The model is able to accurately reproduce both the array-level statistics (see Fig.\ \ref{figure4}\textbf{b}) and individual device behavior (see Fig.\ \ref{figure4}\textbf{c}) observed over the duration of the experiment. Accurate modeling of all the complex dependencies of noise and drift as a function of time and conductance state was found to be very critical in being able to reproduce the experimental evolution of the accuracy on ResNet. 

The resulting accuracy on CIFAR-10 over time is shown in Fig.\ \ref{figure4}\textbf{d}. The test accuracy measured 25 seconds after programming is $93.75\%$, which is very similar to the result obtained in Fig.\ \ref{figure3}\textbf{c}. However, if nothing is done to compensate for conductance drift, the accuracy quickly decreases down to $10\%$ (random guessing) within approximately 1000 seconds. This is because the magnitude of the PCM weights gradually reduces over time due to drift and this prevents the activations from properly propagating throughout the network. A simple global scaling calibration procedure can be used to compensate for the effect of drift on the matrix-vector multiplications performed with PCM crossbars. As proposed in Ref.\ \onlinecite{legalloTED2018}, the summed current of a subset of the columns in the array can be periodically read over time at a constant voltage. The resulting total current is then divided by the summed current of the same columns but read at time $t_0$. This results in a single scaling factor that can be applied to the output of the entire crossbar in order to compensate for a global conductance shift (see Methods and Supplementary Fig.\ 4). Since this factor can be combined with the batch normalization parameters, it does not incur any additional overhead when performing inference. This simple global drift compensation (GDC) procedure was implemented for every layer before carrying out inference on the test set, and the results are shown in Fig.\ \ref{figure4}\textbf{d}. It can be seen that GDC allows the retention of a test accuracy above $92.6\%$ for 1 day on the PCM chip, and effectively prevents the effect of global weight decay over time as illustrated in Supplementary Fig.\ 4. A good agreement of the accuracy evolution between model and experiment is obtained, hence validating its use for extrapolating results over a longer period of time and for assessing the accuracy of larger networks that cannot fit on our current hardware. 

\subsection{Adapting batch normalization statistics to improve the accuracy retention}\label{sec:AdaBS}

Although GDC can compensate for a global conductance shift across the array, it cannot mitigate the effect of $1/f$ noise and drift variability across devices. From the model, we observe that $1/f$ noise is responsible for the random accuracy fluctuations, whereas drift variability and its dependence on the target conductance state cause the monotonous accuracy decrease over time (see Supplementary Fig.\ 5). In order to improve the accuracy retention further, we propose to leverage the batch normalization parameters to correct the activation distributions during inference such that their mean and variance match those that were optimally learned during training. During inference, batch normalization is performed by normalizing the preactivations by their corresponding running mean $\mu$ and variance $\sigma^2$ computed during training. Then, scale and shift factors ($\gamma$ and $\beta$) that were learned through backpropagation are applied to the normalized preactivations. Since $\gamma$ and $\beta$ are learnable parameters, it is not desirable to change them since it would require retraining the model on the PCM devices. However, updating $\mu$ and $\sigma^2$ is more intuitive, since the mean and variance of the preactivations are affected by noise and drift. Leveraging this idea, we introduce a new compensation technique called adaptive batch normalization statistics update (AdaBS), which improves the accuracy retention beyond GDC at the cost of additional computations during the calibration phase. 

\begin{figure*}
\includegraphics[scale=0.95]{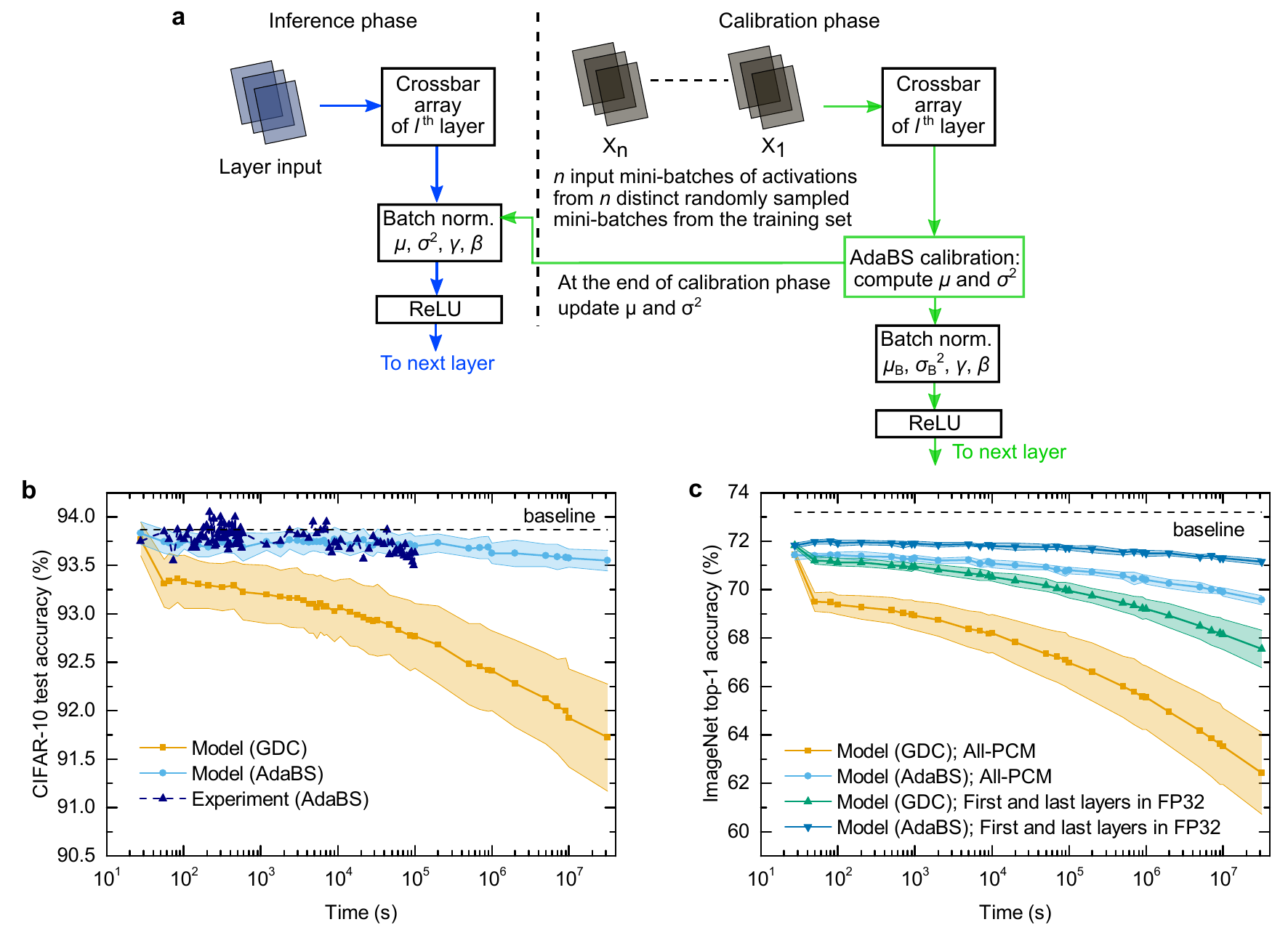}
\caption{\textbf{a}, The AdaBS calibration procedure consists in updating the running mean $\mu$ and variance $\sigma^2$ parameters of the batch normalization performed in the digital unit of the in-memory computing hardware. The calibration is performed periodically when the device is idle, and after calibration the values of $\mu$ and $\sigma^2$ of every layer are updated for subsequent inference. Note that during the calibration phase, batch normalization is performed using the mini-batch mean $\mu_B$ and variance $\sigma_B^2$ instead of $\mu$ and $\sigma^2$ (see Methods and Supplementary Note 3). \textbf{b}, Test accuracy of ResNet-32 on CIFAR-10 with GDC and AdaBS using the PCM model, along with experimental test accuracy obtained by applying AdaBS on ResNet-32 with the measured weights from the PCM chip. The filled areas from the PCM model correspond to one standard deviation over 25 inference runs. \textbf{c} Top-1 accuracy of ResNet-34 on ImageNet with GDC and AdaBS computed using the PCM model. Implementations using PCM synapses for all layers as well as first and last layers in digital FP32, and PCM synapses for all other layers, are shown. In the latter, no noise is applied on the first and last layers during training. The filled areas correspond to one standard deviation over 10 inference runs. } \label{figure5}
\end{figure*}

As described in Fig.\ \ref{figure5}\textbf{a}, the calibration phase consists in sending multiple mini-batches from a set of calibration images that come from the same distribution than the images seen during inference. In this study, we use the images from the training dataset as calibration images. The running mean and variance of preactivations are computed across the entire calibration dataset. The new values of $\mu$ and $\sigma^2$ computed during calibration are then used for subsequent inference. The main advantage of this technique is that it does not incur additional digital computations nor weight programming during inference, since we are only updating the batch normalization parameters $\mu$ and $\sigma^2$ when the calibration is performed. However, injecting the entire training dataset to compute $\mu$ and $\sigma^2$ in the calibration phase would bring significant overhead. When reducing the amount of injected images, the number of updates of the running statistics becomes smaller, and if the momentum used for computing $\mu$ and $\sigma^2$ is not properly tuned to account for this, the network accuracy heavily decreases. To tackle this issue, we developed a procedure to obtain the optimal momentum as a function of the number of mini-batches used for calibration (see Methods and Supplementary Note 3). With this method, we were able to reduce the number of calibration images down to $5.2\%$ of the CIFAR-10 training dataset (2,600 images) without affecting the accuracy. With that number of images, the overhead in terms of digital computations of the AdaBS calibration is about 52\% of performing batch normalization during inference on the whole CIFAR-10 test set (see Supplementary Note 3). It may appear cumbersome to send so many images to the device to perform the calibration, however since it is only performed periodically over time when the device is idle and not every time an image is inferred by the network, the high calibration cost can be amortized. The calibration overhead can be further reduced by using more efficient variants of batch normalization such as the $L^1$-norm version (see Supplementary Note 3). Moreover, although we used AdaBS (and GDC) to compensate solely for the drift of the PCM devices, the same procedure can be applied to mitigate conductance changes due to ambient temperature variations, a critical issue for any analog in-memory computing hardware. The resulting accuracy when performing AdaBS on ResNet-32 with hardware weights before carrying out inference on the test set is shown in Fig.\ \ref{figure5}\textbf{b}. AdaBS allows to retain a test accuracy above $93.5\%$ over one day, an improvement of $0.9\%$ compared with GDC. This improvement becomes $1.8\%$ for one year when extrapolating the results using the PCM model. 

We also applied AdaBS on the ImageNet classification task with ResNet-34, trained with $\eta_{tr} = 3.8\%$, using the PCM model to simulate the weight evolution for one year. By applying the same AdaBS method as for CIFAR-10 using only 0.1\% of the ImageNet training dataset for calibration (1300 images), the accuracy after one year is increased by 7\% compared with GDC when all layers are implemented with PCM synapses (see Fig.\ \ref{figure5}\textbf{c}). When the first and last layers are implemented in digital FP32, the initial accuracy increases to $71.9\%$ and the retention is significantly improved. This technique, combined with AdaBS, allows the retention of an accuracy above $71\%$ for one year. Drawbacks in efficiency when performing inference on hardware in this way have to be mentioned, but they stay limited given the small number of parameters and input size of the first and last layers \footnote{The first layer's input is a large $224 \times 224$ image, but it has only 3 channels. For the last layer, the input is flattened to a single 512-dimensional vector (assuming a batch size of 1). The first and last layers contain less than 3\% of the network weights. }.

\section{Discussion}\label{sec:discussion}

Combined together, the strategies developed in this study allow us to achieve the highest accuracies reported so far with analog resistive memory on the CIFAR-10 and ImageNet benchmarks with residual networks close to their original implementation \cite{Y2016heCVPR}. Although there is still room for improvement especially on ImageNet, those accuracies are already comparable or higher than those reported on ternary weight networks \cite{li2016ternary}, for example $71.6\%$ top-1 accuracy of ResNet-34 on ImageNet with first layer in FP32 \cite{venkatesh2017}. Importantly, the accuracies we report are achieved with just a single nanoscale PCM device encoding the absolute value of a weight. A common approach that could improve the accuracy further is to use multiple devices to encode different bits of a weight \cite{shafiee2016,tsai2019}, at the expense of area and energy penalty, and additional support required by the peripheral circuitry. Aligned with previous observations \cite{Rekhi2019,McKinstry2018}, we notice that retraining ResNet with additive noise results mainly in adapting the batch normalization parameters, whereas the weights stay close to the full-precision weights trained without noise. Hence, retraining by injecting noise from a pretrained baseline network rather than from scratch is very effective since the network recovers high accuracy very quickly, especially for ImageNet. Although our experiments are not done on a fully-integrated chip that supports all functions of deep learning inference, the most critical effects of array-level variability, noise, and drift, are fully accounted for because each weight of the network is programmed on individual PCM devices of our array. Aspects of a fully-integrated chip that are not entirely captured in our experiments such as IR drop and additional circuit nonidealities such as offsets and noise have been studied in previous works and could be mitigated by additional retraining methods \cite{Y2019moonTVLSI,liu2015}. Additional errors due to quantization coming from the crossbar data converters are analyzed further below. 

There exist many different methods of training a neural network with noise that aim to improve the resilience of the model to analog mixed-signal hardware. These include injecting additive noise on the inputs of every layer \cite{Y2019moonTVLSI}, on the preactivations \cite{klachko2019,Rekhi2019}, or just adding noise on the input data \cite{bishop1995}. Moreover, injecting multiplicative Gaussian noise to the weights \cite{murray1994} ($\sigma_{\delta W_{tr},ij}^{l} \propto \vert W_{ij}^{l} \vert$) is also defensible regarding the observed noise on the hardware. We analyzed the four aforementioned methods, attempting to reach the same accuracy demonstrated previously after weight transfer to PCM devices, to identify their possible benefits and drawbacks (see Supplementary Note 4). 
We found that it is possible to adjust the training procedure of all four methods to achieve a similar accuracy on CIFAR-10 after transferring the weights to PCM synapses. Somewhat surprisingly, even adding noise on the input data during training, which is just a simple form of data augmentation, leads to a model which is more resilient to weight perturbations during inference. This shows that it is not necessary to train a model with very complicated noise models that imitate the observed hardware noise precisely. As long as the data propagated through the network is corrupted by a Gaussian noise of the right magnitude, the model is expected to be robust to mapping on PCM devices. 
However, all four methods require one or multiple noise scaling factor hyperparameters to tune in order to reach satisfactory accuracy after transfer to PCM. In contrast, our proposed methodology estimates the additive noise to inject on the weights, $\eta_{tr}$, from a simple hardware characterization, avoiding any hyperparameter search for noise scaling factors. The value of $\eta_{tr}$ does not have to be very precise either, because there is a range of values that lead to similar accuracy after transfer to PCM (see Supplementary Fig.\ 3). 
Moreover, we found that injecting noise on weights achieves better accuracy retention over time (see Supplementary Note 4), which suggests that weight noise mimics the behavior of the PCM hardware better. 

A critical issue for in-memory computing hardware is the need for digital-to-analog (analog-to-digital) conversion every time data goes in (out) of the crossbar arrays. These data conversions lead to quantization of the activations and preactivations, respectively, which introduce additional errors in the forward propagation. Based on a recent ADC survey \cite{Rekhi2019}, 8-bit data conversion is a good tradeoff between precision and energy consumption. Hence, we analyzed the effect of quantizing the input and output of every layer of ResNet-32 and ResNet-34 to 8-bit on the inference accuracy. We set the input/output quantization ranges to the $99.995$-th percentile of the activation/preactivation distributions that are obtained when forward propagating 10k randomly sampled images from the training dataset through the baseline network. As shown in Supplementary Fig.\ 6, even though the 8-bit quantization is not included in our training algorithm, the quantization has a minimal effect on the mean accuracy of ResNet-32 on CIFAR-10 ($<0.05\%$ drop) and ResNet-34 on ImageNet ($<0.15\%$ drop) after weight transfer to PCM synapses. The accuracy evolution over time, retaining the same quantization ranges, does not degrade significantly further and stays well within one standard deviation of that obtained without quantization. The small accuracy deviations could be potentially overcome by including the quantization in the retraining process, which will likely be necessary if less than 8-bit resolution is desired for higher energy efficiency.

Although a computational memory accelerates the matrix-vector multiplication operations in a DNN, communicating activations between computational memory cores executing different layers can become a bottleneck. This bottleneck depends upon two factors, (i) the way different layers are connected to each other and (ii) the latency of the hardware implementation to transfer activations from one core to another. Designing optimal interconnectivity between the cores for state-of-the-art deep CNNs is an open research problem. Indeed, having the network weights stationary during execution in a computational memory puts limits on what portion of the computation can be forwarded to different cores and what cannot. This ultimately results in long-established hardware communication fabrics being ill-fit for the task. One topology for communication fabrics that is well-suited for computational memory is proposed by Dazzi et al.\ \cite{dazzi2019}. It is based on a 5 parallel prism (5PP) graph topology and facilitates inter-layer pipelined execution of CNNs \cite{shafiee2016}. The proposed 5PP topology allows the mapping of all the primary connectivities of state-of-the-art neural networks, including ResNet, DenseNet and Inception-style networks \cite{dazzi2019}. 
As discussed in Ref. \onlinecite{dazzi2019}, the ResNet-32 implementation with 5PP can result in potentially $2 \times$ improvement in pipeline stage latency with similar bandwidth requirements compared with a standard 2D-mesh. Assuming 8-bit activations, communication links with data rate of 5Gbps \cite{sacco2017}, and crossbar computational cycle time of \unit[100]{ns}, a single image inference latency of \unit[52]{$\mu$s} and frame rate of $38,600$ frames per second (FPS) for ResNet-32 on CIFAR-10 is estimated. As an approximate comparison, YodaNN \cite{andri2017}, a digital DNN inference accelerator for binary weight networks with ultra-low power budget, achieves 434.8 FPS in high throughput mode for a 9-layer CNN (BinaryConnect \cite{Y2015courbariauxANIPS}) on CIFAR-10. Although not a direct comparison, the proposed topology and pipelined execution of ResNet-32 could result in $88 \times$ speedup, with a deeper network than the digital solution. 

In summary, we introduced strategies for training ResNet-type CNNs for deployment on analog in-memory computing hardware, as well as improving the accuracy retention on such hardware. We proposed to inject noise to the synaptic weights which is proportional to the combined read and write conductance noise of the hardware during the forward pass of training. This approach combined with judicious weight initialization, clipping, and learning rate scheduling, allowed us to achieve an accuracy of 93.7\% on the CIFAR-10 dataset and a top-1 accuracy on the ImageNet benchmark of 71.6\% after mapping the trained weights to analog PCM synapses. Our methods introduce only a single additional hyperparameter during training, the weight clip scale $\alpha$, since the magnitude of the injected noise can be easily deduced from a one-time hardware characterization. After programming the trained weights of ResNet-32 on 723,444 PCM devices of a prototype chip, the accuracy computed from the measured hardware weights stayed above $92.6\%$ over a period of 1 day, which is to the best of our knowledge the highest accuracy experimentally reported to-date on the CIFAR-10 dataset by any analog resistive memory hardware. A global scaling procedure was used to compensate for the conductance drift of the PCM devices, which was found to be critical in improving the accuracy retention. However, global scaling could not mitigate the effect of $1/f$ noise and drift variability across devices, which led to accuracy fluctuations and monotonous accuracy decrease over time, respectively. Periodically calibrating the batch normalization parameters before inference allowed to alleviate those issues at the cost of additional digital computations, increasing the 1-day accuracy to $93.5\%$ on hardware. These results demonstrate the feasibility of realizing accurate inference on complex DNNs through analog in-memory computing using existing PCM devices. 

\clearpage
\section*{Methods}

\subsection{Experiments on PCM hardware platform}

The experimental platform is built around a prototype PCM chip that comprises 3 million PCM devices. The PCM array is organized as a matrix of word lines (WL) and bit lines (BL). In addition to the PCM devices, the prototype chip integrates the circuitry for device addressing and for write and read operations. The PCM chip is interfaced to a hardware platform comprising two field programmable gate array (FPGA) boards and an analog-front-end (AFE) board. The AFE board provides the power supplies as well as the voltage and current reference sources for the PCM chip. The FPGA boards are used to implement overall system control and data management as well as the interface with the data processing unit. The experimental platform is operated from a host computer, and a Matlab environment is used to coordinate the experiments. The PCM devices were integrated into the chip in 90-nm CMOS technology using the key-hole process described in Ref. \onlinecite{breitwischVLSI2007}. The phase-change material is doped Ge$_2$Sb$_2$Te$_5$. The bottom electrode has a radius of $\sim 20$ nm and a length of $\sim 65$ nm. The phase-change material is $\sim 100$ nm thick and extends to the top electrode, whose radius is $\sim 100$ nm. All experiments performed in this work were done on an array containing 1 million devices accessed via transistors, which is organized as a matrix of 512 WL and 2048 BL.

A PCM device is selected by serially addressing a WL and a BL. To read a PCM device, the selected BL is biased to a constant voltage (\unit[$300$]{mV}) by a voltage regulator via a voltage generated off chip. The sensed current is integrated by a capacitor, and the resulting voltage is then digitized by the on-chip 8-bit cyclic analog-to-digital converter (ADC). The total duration of applying the read pulse and converting the data with the ADC is \unit[$1$]{$\mu$s}. The readout characteristic is calibrated via on-chip reference polysilicon resistors. To program a PCM device, a voltage generated off chip is converted on chip into a programming current. This current is then mirrored into the selected BL for the desired duration of the programming pulse. Iterative programming involving a sequence of program-and-verify steps is used to program the PCM devices to the desired conductance values \cite{papandreouISCAS2011}. The devices are initialized to a high-conductance state via a staircase-pulse sequence. The sequence starts with a RESET pulse of amplitude \unit[450]{$\mu$A} and width \unit[50]{ns}, followed by 6 pulses of amplitude decreasing regularly from \unit[160]{$\mu$A} to \unit[60]{$\mu$A} and with a constant width of \unit[1000]{ns}. After initialization, each device is set to a desired conductance value through a program-and-verify scheme. The conductance of all devices in the array is read 5 times consecutively at a voltage of \unit[0.3]{V}, and the mean conductance of these reads is used for verification. 
If the read conductance of a specific device does not fall within \unit[0.25]{$\mu$S} from its target conductance, it receives a programming pulse where the pulse amplitude is incremented or decremented proportionally to the difference between the read and target conductance. The pulse amplitude ranges between \unit[80]{$\mu$A} and  \unit[400]{$\mu$A}. This program-and-verify scheme is repeated for a maximum of 55 iterations.

In the hardware/software inference experiments, the analog conductance values of the PCM devices encoding the network weights, $G_{ij}^{l,+}$ and $G_{ij}^{l,-}$, are serially read individually with the 8-bit on-chip ADC at predefined timestamps spaced over a period of one day. The read conductance values at every timestamp are reported to a TensorFlow-based software.  This software performs the forward propagation of the CIFAR-10 test set on the weights read from hardware and computes the resulting classification accuracy. The drift compensation techniques, GDC and AdaBS, are performed entirely in software at every timestamp based on the conductance values read from hardware. 

\subsection{PCM-based deep learning inference simulator}

We developed a simulation framework to test the efficacy of DNN inference using PCM devices. We chose Google's TensorFlow\citep{TF} deep learning framework for the simulator development. The large library of algorithms in TensorFlow enables us to use native implementation of required activation functions and batch normalization. Moreover, any regular TensorFlow code of a DNN can be easily ported to our simulator. As shown in Supplementary Fig.\ 7, custom made TensorFlow operations are implemented that generate PCM conductance values from the behavioral model of hardware PCM devices that was developed (see Supplementary Note 2). All the nonidealities including conductance range, programming noise, read noise, and conductance drift are implemented in TensorFlow following the equations shown in Supplementary Note 2. The simulator can also take the PCM conductance data measured from hardware as input, in order to perform inference on the hardware data. Data converters that simulate digital quantization of data at the input and output of crossbars are also implemented with tunable quantization ranges and precision. In this study, the data converters were turned off for all simulations except those presented in Supplementary Fig.\ 6. The drift correction techniques are implemented post quantization of the crossbar output. 

\subsection{Training implementation of ResNet-32 on CIFAR-10}

ResNet-32 has 31 convolution layers with $3\times3$ kernels, 2 convolution layers with $1\times1$ kernels, and a final fully-connected layer. The network contains 361,722 synaptic weights.  It consists of 3 different ResNet blocks with 10 $3\times3$ kernels each.  After the first convolution layer, there is a unity residual feed forward connection after every two convolution layers, except the $1\times1$ residual convolution connection to make output channels compatible between two layers.  Each convolution layer is followed by batch normalization \cite{batchnorm}. ReLU activation is used after every batch normalization except in case of residual connections, where the ReLU activation is computed after summation.  The output of the last convolution layer is then downsampled using global average pooling \cite{GAP}, which is followed by a single fully-connected layer. For the last fully-connected layer, no batch normalization is performed. The architecture of ResNet-32 used in this study is a slightly modified version of the original implementation \cite{Y2016heCVPR} with fewer input and output channels in ResNet blocks 2 and 3. This network is trained on the well-known CIFAR-10 classification dataset \cite{cifar10}. It has $32\times32$ pixels RGB images that belong to one of the 10 classes.  

The network is trained on the 50,000 images of the training set, and evaluation is performed on the 10,000 images of the test set.  The training is performed using stochastic gradient descent with a momentum of 0.9. The network objective is categorical cross entropy function over 10 classes of the input image. Learning rate scheduling is performed to reduce learning rate by 90\% at every 50$^{th}$ training epoch. The initial learning rate for the baseline network is 0.1 and training converges in 200 epochs with a mini-batch size of 128.  Weights of all convolution and fully connected layers of the baseline network are initialized using He Normal \cite{he_init} initialization. The baseline network is retrained by injecting Gaussian noise for up to 150 epochs with weight clip scale $\alpha = 2$. We preprocess the training images by randomly cropping a $32\times32$ patch after padding 2 pixels along the height and width of the image. We also apply a random horizontal flip on the images from the train set. Additionally, we apply cutout\cite{cutout} on the training set images. For both training and test set, we apply channel wise normalization for 0 mean and unit standard deviation.

\subsection{Training implementation of ResNet-34 on ImageNet}

The architecture of the ResNet-34 network for ImageNet classification is derived from Ref. \onlinecite{Y2016heCVPR}. It has 32 convolution layers with $3\times3$ kernels, 3 convolution layers with $1\times1$ kernels, a first convolution layer with $7\times7$ kernels and a final fully-connected layer. The network has 21,797,672 synaptic weights. The first convolution layer downsamples the input by using a stride of 2 pixels, followed by a maxpooling layer with kernel size of $3 \times 3$ and stride of 2 to downsample the feature maps to the resolution of $56 \times 56$ pixels. Each residual connection with $1 \times 1$ convolution and first layer of ResNet blocks 2, 3, 4 downsample the input by using a stride of 2 pixels. A global average pooling layer before the final fully-connected layer downsamples the $7 \times 7$ input to $1 \times 1$ resolution. The final fully-connected layer computes the output prediction corresponding to 1,000 classes.

We trained ResNet-34 on the ImageNet\cite{imagenet} dataset. The ImageNet dataset has 1.3M images in the training set and 50k images in the test set. Images in the ImageNet dataset are preprocessed by following the same preprocessing steps as that of the Pytorch baseline model. Training images are randomly cropped to a $224 \times 224$ patch and then random horizontal flip is applied on the images. Channel wise normalization is performed on the images in both training and test set for 0 mean and unit standard deviation. Only for the test set, images are first resized to $256 \times 256$ using bilinear interpolation method and then a center crop is performed to obtain the $224 \times 224$ image patch. 

The network objective function is softmax cross entropy on network output and corresponding 1,000 labels. The network objective is minimized using the stochastic gradient descent algorithm with a momentum of 0.9. We obtained our baseline network architecture and its parameters from the Pytorch model zoo\cite{pytorch_model_zoo}. We use this network to perform additive noise training by injecting Gaussian noise for a total of 10 training epochs. In contrast to ResNet-32 on CIFAR-10, no learning rate scheduling was performed since the network was trained only for 10 epochs with additive noise. We use mini-batch size of 400 and learning rate of 0.001 for the additive noise training simulations. We also use L2 weight decay of 0.0001 and weight clip scale of $\alpha = 2.5$ for the additive noise training.

\subsection{Global drift compensation (GDC) method}

The GDC calibration phase consists of computing the summed current of $L$ columns in each array encoding a network layer (see Supplementary Fig.\ 4). Those $L$ columns contain devices initially programmed to known conductance values $G_{mn}(t_0)$. By reading those column currents, $I_m$, periodically with applied voltage $V_{cal}$ on all rows, we can compensate for a global conductance shift in the array during inference. When input data is processed by the crossbar during inference, the crossbar output can be scaled by $1/ \hat{\alpha}$, where
\[ \hat{\alpha} = \frac{\sum_{m=1}^L I_m}{V_{cal} \sum_{n=1}^N \sum_{m=1}^L G_{mn}(t_0)}. \] 
This procedure is especially simple because $L$ can be chosen to be small, enough to get sufficient statistics. Moreover, $\hat{\alpha}$ is computed from the device data itself, without resorting to any assumption on how the conductance changes nor requiring extra timing information. The term $V_{cal} \sum_{n=1}^N \sum_{m=1}^L G_{mn}(t_0)$ needs to be computed only once, stored in the digital memory of the chip, and is reused for all calibrations. Reading the subset of $L$ columns of the crossbar can be done while the PCM array is idle, i.e., when there are no incoming images to be processed by the device. Performing the $L$ current summations can be implemented either with on-chip digital circuitry or in the control unit of the chip. At the end of the calibration phase, $1/ \hat{\alpha}$ is computed and stored locally in digital unit of the crossbar. The output scaling by $1/ \hat{\alpha}$ during inference can be combined with batch normalization because it is a linear operation. In our experiments, the calibration procedure was performed using all columns of each layer (e.g. $L$ is equal to two times number of output channels) every time before inference is performed on the whole test set. 

\subsection{Adaptive batch normalization statistics update (AdaBS) technique}

Batch normalization is performed differently in the training and inference phases of a DNN. During the training of a DNN, batch normalization normalizes the input to zero mean and unit variance by computing the mean ($\mu_{B}$) and variance ($\sigma_{B}^{2}$) over the mini-batch of $m$ images
\begin{eqnarray}
\mu_B &=& \frac{1}{m} \sum_{i=1}^m x_i \\
\sigma^2_B &=& \frac{1}{m} \sum_{i=1}^m (x_i - \mu_B)^2. 
\end{eqnarray}
The normalized input is then scaled and shifted by $\gamma$ and $\beta$. During the training phase, $\gamma$ and $\beta$ are learned through backpropagation. In parallel, a global running mean ($\mu$) and variance ($\sigma^{2}$) are computed by exponentially averaging $\mu_{B}$ and $\sigma_{B}^{2}$ respectively, over all the training batches 
\begin{eqnarray}
\mu &=& p \cdot \mu + (1-p)\cdot \mu_B \\
\sigma^2 &=& p \cdot \sigma^2 + (1-p)\cdot \sigma^2_B, 
\end{eqnarray}
where $p$ is the momentum. After training, the estimates of the global mean and variance $\mu$ and $\sigma^{2}$ are then used during the inference phase. When performing forward propagation during inference, the batch normalization coefficients $\mu$, $\sigma^{2}$, $\gamma$, and $\beta$ are used for normalization, scale, and shift. 

The calibration phase of AdaBS consists in recomputing and updating $\mu$ and $\sigma^{2}$ for every layer where batch normalization is present. We recompute $\mu$ and $\sigma^{2}$ by feeding a randomly sampled set of mini-batches from the training dataset. In recomputing $\mu$ and $\sigma^{2}$, hyper-parameters such as mini-batch size ($m$) and momentum ($p$) need to be carefully tuned to achieve the best network accuracy.
	
For AdaBS calibration, we observed that using an optimal value of the momentum is necessary to achieve good inference accuracy evolution over time. For this, we have developed an algorithm to estimate the optimal value of momentum by an empirical analysis, which is explained in Supplementary Note 3. Based on this analysis, the formula we used to compute the optimal momentum as a function of the number of injected mini-batches $n$ is
\begin{equation}
\label{EQ:optimal_momentum}
p = 0.015^{(1/n)}. 
\end{equation}
Using Eq.\ \eqref{EQ:optimal_momentum} to compute the momentum, we found that with a fixed mini-batch size of $m=200$ images, it is sufficient to inject $n=13$ mini-batches for the AdaBS calibration of the ResNet-32 network, that is approximately $5\%$ of the CIFAR-10 training set (2,600 images). The sensitivity of the accuracy to the number of images used for AdaBS calibration is shown in Supplementary Note 3. For ResNet-34 on ImageNet, we used mini-batch size of $m=50$ and $n=26$ mini-batches, that is  $0.1\%$ of the ImageNet training set (1,300 images). In the experiments presented in Fig.\ \ref{figure5}, AdaBS calibration was performed for every layer before performing inference on the test set, except the last layer because it does not have batch normalization.

\section*{Acknowledgments}
We thank O. Hilliges for discussions, and our colleagues at IBM TJ Watson Research Center, in particular M. BrightSky, for help with fabricating the PCM
prototype chip used in this work. This work was partially funded by the European Research Council (ERC) under the European Union's Horizon 2020
research and innovation programme (grant agreement number 682675).

\section*{Author Contributions}
V.J., M.L., S.H., C.P. and A.S. conceived the training methodology. V.J., M.L., S.H., I.B. and A.S. conceived the drift correction techniques. V.J. and S.H. performed the software training and inference simulations under the guidance of M.L.. I.B. performed the PCM hardware experiments with the support of V.J.. S.R.N. and V.J. developed the PCM model. V.J. and C.P. developed the PCM deep learning inference TensorFlow-based software. M.D. provided critical in-memory computing hardware insights and performed the ResNet-32 performance estimation. M.L. wrote the manuscript with input from all authors. M.L., A.S., B.R. and E.E. supervised the project.

\clearpage
\section*{References}
\bibliography{MPDL}

\end{document}